\input epsf
\documentstyle[12pt]{article}
\begin{document}
\begin{titlepage}
\setcounter{page}{0}
\rightline{PACS 13.25.Hw, 12.15.Ji}
%\rightline{Preprint IC/96/193}
\vspace{1cm}
\begin{center}

{\Large  CP-Violation in b Quark Radiative Inclusive Decays}\\
\vspace{1cm}
{\large H. M. Asatrian$^a$} \\
{\em ~~ International Centre for Theoretical Physics, Strada Costiera
11, P.O. Box 586, 34100 Trieste, Italy\\
e-mail: hrachia@moon.yerphi.am}\\
\vspace{0.4cm}
{\large G. K. Yeghiyan$^a$}\\
{\em Deutches Electronen Synchrotron DESY, Hamburg, Germany}\\
\vspace{0.4cm}
{\large A. N. Ioannissian$^a$} \\
{\em Dept. of Physics, Technion - Israel Inst. of Tech., Haifa,
Israel}\\
\end{center}
\vspace{1cm}

\centerline{\bf{Abstract}}
\vspace{0.3cm}
The direct CP-violations in  $SU(2)_{L}\times SU(2)_{R} \times U(1)$
model and two-Higgs doublet extension of the standard model for
$b \rightarrow d\gamma$ and $b \rightarrow s\gamma$ decays  are
investigated. The calculated value of CP-asymmetry for these
two models and  for $b \rightarrow d\gamma$ and $b \rightarrow s\gamma$
decays for the wide range of parameters may  exceed the value,
predicted by the standard model and has a sign opposite to that of the latter.

\vfill
\begin{center}
{\em $^a$ Permanent address: Yerevan Physics Institute, 2 Alikhanyan Br.,
375036, Yerevan, Armenia}\\
\end{center}
\end{titlepage}
\newpage
The investigation of rare B-meson decays can give
an important information on new physics in the TeV region.
The observation of direct CP asymmetry in B-meson decays will help
to understand the CP breaking phenomenon.\\
The first experimental evidence for the exclusive
$\bar{B} \rightarrow K^{*} \gamma$ decay has been obtained at  CLEO  \cite{1}. More recently, the branching ratio of the inclusive 
$B \to X_s \gamma$ decay was measured \cite{2}. \\
The $b \rightarrow s \gamma$ decay has been investigated theoretically
for the standard model and its  extensions in \cite{3}-\cite{12}.
CP-violation in $B-\bar{B}$ system in $SU(2)_{L}\times SU(2)_{R} \times
U(1)$ model was considered in \cite{13}. The problem of CP
asymmetry for $b \rightarrow s \gamma$ decay for standard model and its
extensions was investigated in \cite {14,15,16}.\\
Although the expected decay rate for $b \rightarrow d \gamma$ decay
is about 10-20 times smaller than for the $b \rightarrow s \gamma$ decay,
the CP asymmetry for the first decay can be about 10 times larger
\cite{14,17}.\\
The aim of the present paper is to consider the  direct CP decay asymmetry  for the $b \rightarrow d \gamma$ and $b \rightarrow s \gamma$ decays for
$SU(2)_L \times SU(2)_R \times U(1)$
model and two Higgs doublet extension of the standard model.\\
%The obtained value of CP-asymmetry for $b \rightarrow d \gamma$
%and for $b \rightarrow s \gamma$ decays  for the reasonable range of
%parameters of models (the right W-boson mass, the charged scalar
%particle mass, the ratio of  two Higgs doublets vacuum expectation
%values, phases and mixing angles) can have a sign opposite to that in
%the standard model. The absolute value of the asymmetry may far exceed
%the standard model predictions.\\
In $SU(2)_{L}\times SU(2)_{R} \times U(1)$ model the $b \rightarrow d \gamma$
decay amplitude arises due to the interaction of quark charged weak current with the "left" and "right" W-bosons and charged Higgs field. This
interaction has the following form \cite{16}:
\begin{eqnarray}
L^{ch}&=&\frac{1}{\sqrt{2}}\left(\bar{u},\bar{c},\bar{t}\right)\left[
\hat{W_1}^+\left[-g_L\cos\beta K^L P_--g_R\sin\beta  e^{i\delta }K^RP_+
\right]\right.+\nonumber \\
&+&\varphi ^+\frac{g_L }{\sqrt{2}M_{W_L}}\left[ \left(-\tan2\theta K^LM_d+
e^{i\delta }\frac{1}{\cos2\theta}M_uK_R\right)P_++\right.\\
&+&\left.\left.\left( \tan2\theta                                      %(1)
 M_uK_L-e^{i\delta }\frac{1}{\cos2\theta}K^RM_d\right)
P_-\right]\right]\left(\begin{array}{l}
d\\s\\b \end{array}\right)\nonumber,
\end{eqnarray}
where $W_1$ is the "light" (predominantly
left-handed) charged gauge boson
and $\beta$ is the mixing angle between left and right W - bosons,
\begin{equation}
\tan2\beta  =2\sin2\theta
\frac{g_R}{g_L}
\frac{M^2_{W_L}}{M^2_{W_R}}
\hspace{0.4cm} \tan\theta=-\frac{k'}{k},                               %(2)
\end{equation}
$K^{L}$ and $K^{R}$ are CKM mixing matrices
for left and right charged currents respectively,
$P_{\pm}=(1\pm \gamma_{5})/2$, $M_u$ and $M_d$ are diagonal
mass matrices for quarks with Q=2/3 and Q=-1/3 charges respectively.
The matrices $K^{L}$ and $K^{R}$ can be expressed in a form, where $K^{L}$
has only one complex phase and $K^{R}$ has five complex phases
\cite{18}. Phase $\delta$ in (2) takes his origin  from the vacuum
expectation value of Higgs field $\Phi$, connected with the
$SU(2)_L \times U(1)$ symmetry breaking:
$\Phi=
\left( \begin{array}{cc}
k & 0\\
0 & -e^{i\delta} k'\\
\end{array} \right)$.
In (1) the term connected with the interaction with heavy
(predominantly right) W- boson is omitted, since it is not relevant
for $b \rightarrow d \gamma$ decay.\\
The additional (compared with the standard model) phase factor
$exp(i\delta)$ in (1) leads to the existence of the new CP violation effects in
$SU(2)_L \times SU(2)_R \times U(1)$
model.\\
We define the direct CP asymmetry  for $b \rightarrow d\gamma$ decay as
\cite{14}:
\begin{equation}
a_{cp}=\frac{\Gamma(\bar{b} \rightarrow \bar{d}\gamma)-
\Gamma(b \rightarrow d\gamma)}
{\Gamma(\bar{b} \rightarrow \bar{d}\gamma)+
\Gamma(b \rightarrow d\gamma)}                                        %(3)
\end{equation}
The direct CP asymmetry for $b \rightarrow d\gamma$ decay arises only if
the matrix element of decay has an absorptive part,
which arises if the final state strong
interaction effects are taken into account.
In general case the amplitudes of the decays
$\bar{b} \rightarrow \bar{d}\gamma$
and $b \rightarrow d\gamma$ can be expressed in the following form
\cite{14}:
 \begin{eqnarray}
\nonumber
A(\bar{b} \rightarrow \bar{d}\gamma)=\sum_a (A_a^r+iA_a^i) V_a^* \\    %(4)
A(b \rightarrow d\gamma)=\sum_a (A_a^r+iA_a^i) V_a
\end{eqnarray}
where $A_a^r$, $A_a^i$ are real and absorptive parts of amplitudes, $V_a$
are some phases and CKM-type factors, a=1,2,3....
Then CP asymmetry is given by:
\begin{equation}
a_{CP}=\frac{\sum_{a\neq b}(A_a^r A_b^i-A_a^i A_b^r) Im (V_a^* V_b)}
{\sum_{a,b}(A_a^r A_b^r+A_a^i A_b^i) Re(V_a^* V_b)}                     %(5)
\end{equation}
%It is significant, that in the general case the CP asymmetry depends not
%only on imaginary but also on real part of CKM matrix.\\
To take into account QCD- corrections to radiative decays matrix elements
the effective Hamiltonian approach
is used. We follow to \cite{16} and use the results of
\cite{12} for the imaginary part of the amplitude, connected with the $O_2$
operator:
\begin{displaymath}
Imr_2=\frac{16\pi}{81}[-5+(45-3\pi^2+9L+9L^2)z+(-3\pi^2+9L^2)z^2+
(28-12L)z^3]
\end{displaymath}
where $z=(m_c/m_b)^2$, L=lnz.
We take the ratio of c- and b- quark masses  equal to
0.29 \cite{12}, then the ratio of imaginary parts of the amplitudes
connected with $\bar{c}c$ and $\bar{u}u$ intermediate states
is approximately  equal to r=0.145. We obtain the following
expression for effective Hamiltonian of  $b \rightarrow d\gamma$ decay
in $SU(2)_L \times SU(2)_R \times U(1)$ model:
\begin{eqnarray}
\nonumber
H_{b\rightarrow d\gamma} &  =  & - \frac{e}{16\pi^2}
\frac{2G_F}{\sqrt{2}}m_b\{ (K_{td}^{L*}K_{tb}^L A_{d\gamma }^{W_L}
+ e^{i\delta}K_{td}^{L*}K_{tb}^R A_{d\gamma}^{R})O_7^L\\
& + &e^{-i\delta}K_{td}^{R*}K_{tb}^L
A_{d\gamma }^{R}O_7^R
 + i[(\frac{2}{9}(K^{L_*}_{td}K^L_{tb} A^{W_L}_{dg}+
e^{i\delta}K^{L^*}_{td}K^L_{tb}
\frac{K^R_{tb}}{K^L_{tb}}A^R_{dg})\\
\nonumber
& + &\frac{20}{81}(K^{L^*}_{ud}K_{ub}^L                   %(6)
y+rK^{L^*}_{cd}K^L_{cb})c_2 )O_7^L+
\frac{2}{9}e^{-i\delta}K^{L^*}_{td}K^L_{tb}
\frac{K^R_{td}}{K^L_{td}}A^R_{dg}O_7^R]\},
\end {eqnarray}
where
\begin{eqnarray}
O_7^{L,R}=\bar{u}_d \sigma^{\mu\nu}(1\pm \gamma
_5)u_b F_{\mu \nu }                                       %(7)
\end{eqnarray}
and the functions
$A_{d\gamma}^{W_{L}}$,
$A_{d\gamma}^{R}$,
$A_{dg}^{W_{L}}$,
$A_{dg}^{R}$,
which include leading logarithmic strong interaction corrections,
were presented in  \cite{16}.\\
Using (5) and (6), one obtains the following formula for CP asymmetry
for $b\to d \gamma $ decay in $SU(2)_{L}\times SU(2)_{R} \times U(1)$ model:
\begin{eqnarray}
\nonumber
&&a_{cp}(b\to d\gamma)=\frac{2\alpha
_s}{(|C_7^L|^2+|C_7^R|^2) v_t^*v_t}
\left\{({\rm Im} v^*_t v_u+
r{\rm Im}v^*_tv_c)\times \right.\\                                 %(8)
&&\times (A^{W_{L}}_{d\gamma }+
H\cos \alpha  A^R_{d\gamma })\frac{20c_2}{81}-({\rm Re} v^*_t v_u+
r{\rm Re} v^*_t v_c)\times \\
\nonumber
&&\left.\times A^R_{d\gamma }H \sin \alpha
\frac{20c_2}{81}+
\frac{2}{9}H \sin \alpha v^*_t v_t(A_{d\gamma}^{W_{L}}A^R_{dg}-
A^R_{d\gamma }A_{dg}^{W_{L}})\right\} ,
\end{eqnarray}
where
\begin{eqnarray}
\nonumber
&&He^{i\alpha  }\equiv e^{i\delta }\frac{K_{Rtb}}{K_{Ltb}},\quad
C_7^L= A_{d\gamma }^{W_L}+\left(e^{i\delta}\frac{K^R_{tb}}{K^L_{tb}}
\right)
A^R_{d\gamma },\quad C_7^R= e^{-i\delta}\frac{K^{R^*}_{td}}{K^{L^*}_{td}}
A^R_{d\gamma}\\                                  %(9)
&&v_t\equiv K_{Ltd}^*K_{Ltb},\quad \quad v_c\equiv
K_{Lcd}^*K_{Lcb},\quad \quad v_u\equiv K_{Lud}^*K_{Lub}
\end{eqnarray}
The expression for CP
asymmetry in $b\to s \gamma$ (in the limit $m_d$=0, $m_s$=0) decay can be
obtained from (8) by making replacements $v_u\to v_u^{'}$,
$v_c\to v_c^{'}$, $v_t\to v_t^{'}$, where
\begin{equation}
v_t^{'}\equiv K_{Lts}^*K_{Ltb},\quad \quad v_c^{'}\equiv
K_{Lcs}^*K_{Lcb},\quad \quad v_u^{'}\equiv K_{Lus}^*K_{Lub}       %(10)
\end{equation}

We want to stress that in \cite{14,15,16} the approximate value for
r was used: r=0.12. Now we use the correct value r=0.145 \cite{12},
which is essential for the numerical results. \\
Let us proceed to the numerical results. We take $\alpha_s=0.212$,
$c_{2}\simeq 1.1$, $m_t=(175 \pm 9)$GeV, $m_b=4.5GeV$,
$m_b^*\equiv m_b(M_Z)=(3.5\pm 0.5)GeV$ \cite{19,20}.
For CKM matrix parameters we use Wolfenstein parametrization:
\begin{eqnarray}
\nonumber
&&Im(v_t^{*}v_u)=-A^2\lambda^6 \eta
\hspace{1.5cm}
Re(v_t^{*}v_u)=A^2\lambda^6((1-\rho)\rho-\eta^2) \\  %(11)
\nonumber
&&|v_t|^2= A^2\lambda^6((1-\rho)^2+\eta^2)
\hspace{1.5cm}
|v_u|^2= A^2\lambda^6(\rho^2+\eta^2)\\
&&|v_t{'}|^2= A^2\lambda^4((1+\lambda^2\rho)^2+\lambda^8\eta^2)
\hspace{0.5cm}
Im(v_t^{'*}v_u^{'})=A^2\lambda^6 \eta\\
\nonumber
&&Re(v_t^{'*}v_u^{'})=-A^2\lambda^6(\rho+\lambda^2(\rho^2+\eta^2))
\hspace{0.5cm}
|v_u^{'}|^2= A^2\lambda^8(\rho^2+\eta^2)
\end{eqnarray}
For parameters $\lambda$, A, $\rho$, $\eta$ in (11)  we use values given in
\cite{20}.\\
The CP asymmetry depends also on parameters of
$SU(2)_L \times SU(2)_R\times U(1)$ model: $\alpha$, $tan2\theta$,
$M_{W_R}$, $M_{\varphi^+}$, $|K^R_{tb}/K^L_{tb}|$, $|K^R_{td}/K^L_{td}|$.\\
%We want to make clear for which values of the parameters the CP asymmetry
%for considered decays differs significantly from the standard model
%prediction.\\
%To reduce the number of the parameters we will consider in numerical
%calculations that the decay rates of both of the  decays in
%$SU(2)_L \times SU(2)_R\times U(1)$ model
%are close to their  values in the standard model: for
%$b \to s \gamma$ decay the standard model predictions are in
%agreement with experiment and we will consider that it is true for
%$b \to d \gamma$ decay too (for the case (1) (see later) these
%two conditions are identical). Numerically we require that the
%decay rate can differ from the standard model predictions no more
%than $\Delta=10\%$ (for two Higgs doublet model we use also less
%severe condition with $\Delta$=50\%).\\
We will consider the following possibilities
for the ratios $|K^R_{tb}/K^L_{tb}|$, $|K^R_{td(s)}/K^L_{td(s)}|$:\\
(1) $|K^R_{tb}/K^L_{tb}|$= $|K^R_{td(s)}/K^L_{td(s)}|$=1\\
(2) No restrictions on ratios $|K^R_{tb}/K^L_{tb}|$, $|K^R_{td(s)}/K^L_{td(s)}|$, besides those which follow from the unitarity conditions for matrices $K^L$ and $K^R$.\\
The case (1) corresponds to the pseudo-manifest left-right symmetry,
when the absolute values of Kobayashi-Maskawa mixing matrices elements in
left and right sectors ($K^L_{ij}$ and $K^R_{ij}$, i,j=1,2,3
correspondingly) are equal to each other \cite{18}.
The case (2) corresponds to non-manifest left-right symmetry
when $|K^L_{ij}|\neq |K^R_{ij}|$ \cite{18}.\\
It is known that the experiment  is in agreement
with the standard model predictions for  $b \to s\gamma$ decay rate. Following to \cite{16},
we will consider that in
$SU(2)_L\times SU(2)_R \times U(1)$ model the branching of
$b \to s\gamma$ decay can differ from the standard model prediction 
no more than $\Delta=10\%$. As for $b \to d\gamma$ decay rate, there is no experimental restriction for it.   However, if we assume that in $SU(2)_L\times SU(2)_R \times U(1)$ model
the $b \to s\gamma$ decay rate  is the same (with 10\% accuracy)
as in standard model, then for the case (1), the same condition will be
satisfied  for the $b \to d\gamma$ decay rate also.  For the
case (2) we will consider two possibilities:\\
(2a) $b \to d\gamma$ decay rate is equal with accuracy of $\Delta=10\%$
to that in the standard model\\
(2b) $b \to d\gamma$ decay rate is arbitrary.\\
%The condition (1) (or (2)) gives the restriction for $\alpha$.
%As we already have mentioned above, to obtain the significant
%deviation from the standard model predictions the
%contribution of the amplitudes $A^R_{d\gamma}$ and $A^R_{dg}$
%to CP-asymmetry must be compatible with the standard model
%amplitude $A^{W_L}_{s\gamma}$. On the other hand the condition
%(1) or (2) must be satisfied. Both of first are increasing
%when the $tan2\theta$ is increasing.
For case (1) and for a
given $M_{W_R}$, $M_{\varphi^+}$ the decay asymmetry for
$b \to d \gamma$ and  $b \to s \gamma$ decays   depends on
CKM parameters, $\alpha$ and $tan2\theta$. Taking into
account (8) and the equivalence of decay rates (with 10\% accuracy)
in standard model and $SU(2)_L\times SU(2)_R \times U(1)$ model,
it is easy to  understand, that for $\alpha=0$ the absolute value of
decay asymmetry for both of decays can't exceed the standard model value
more than 10\% for all the values $M_{W_R}$, $M_{\varphi^+}$,
$tan2\theta$. The sign of the decay asymmetry will be the same
as in the standard model.
When we have a new source of CP violation, i.e.  $\alpha \neq 0.$, the
terms in (8) proportional to $A^R_{d\gamma}$ and
$A^{W_L}_{d\gamma} A^R_{dg}-A^{R}_{d\gamma} A^{W_L}_{dg}$
contribute to decay asymmetry and one can expect less or more
significant deviations from the standard model predictions.
However, the restriction for decay rate here also plays the important
role. In Fig 1 the $tan2 \theta$ dependence of maximum and
minimum values of $a_{CP}$   for $b \to d \gamma$ and  $b \to s \gamma$ decays for the case (1) are given  for various values of
$M_{W_R}$, $M_{\varphi^+}$. Due to the presence of the terms proportional
to $sin\alpha$ in (8) there is a difference between standard model
predictions ($-a_{CP}(b\to s\gamma)/10^{-3} = 2.9 \div 6.4$,
$a_{CP}(b\to d\gamma)/10^{-2}=3.7 \div 16$), which practically
coincide with the results for $M_{W_R}=M_{\varphi^+}$=50TeV and
predictions of the $SU(2)_L \times SU(2)_R \times U(1)$ model for
$M_{W_R},M_{\varphi^+} \leq 20TeV$, $tan2\theta \geq 1$.  The difference
is most significant for $b\to s\gamma$ decay. The sign of asymmetry
can be different from those in the standard model for both  decays.
To illustrate the $\alpha$ dependence of the decay asymmetry we give in
the Table 1 minimum and maximum values of decay asymmetry for two decays for $tan2\theta=2$, $M_{W_R}=M_{\varphi^+}=10TeV$ and various values of $\alpha$ (for $2.30\leq|\alpha|\leq 3.14$ and
$|\alpha|\leq 1.9$ the condition for decay  rate is not satisfied).
As we have mentioned above, the difference between results for $a_{CP}$
when taking r=0.12 or 0.145 in the expression (8) is non-negligible:
for $b \to s \gamma$ decay it can reach  30\%. For this reason
values of $a_{CP}$ in Fig 1 are lower than those in \cite{16}.\\
Let us now proceed to the case (2). In Fig 2a we give the 
$tan2\theta$ dependence
of $a_{CP}$ for $b\to d\gamma$ decay for case (2a)
and for $b\to s\gamma$ decay for case (2).
It is obvious that for the case (2), when there are no restrictions on
right current mixing matrix (besides unitarity condition), the
decay asymmetry in $SU(2)_L\times SU(2)_R\times U(1)$ model is much more
different from the standard model predictions, than for the case (1).
Indeed, for the case (2) the minimum value of the  asymmetry for 
$b \to d\gamma$ decay can reach (for $M_{W_R}=M_{\varphi^+}=5TeV$)  the value
-0.18, while for the previous case the minimum value of $a_{CP}$ for the 
same values $M_{W_R}$ and $M_{\varphi^+}$ is equal to -0.02.
For $b\to s\gamma$ decay the absolute value of decay asymmetry for the 
same values $M_{W_R}$ and $M_{\varphi^+}$ is 1.5-2 times higher than for the
previous case.\\
We give in Table 2 minimum and maximum values of decay asymmetry
for two decays for $tan2\theta=3$, $M_{W_R}=M_{\varphi^+}=10TeV$ and various values of $\alpha$ (for $\alpha|\leq 1.80$ the conditions for decay  rate is not satisfied). It is clear that for the case (2a) the deviations
from the standard model predictions are more significant and
can take place for a larger parameter space, than for the case (1).\\
In Fig 3 for the case (2b) (when we have no restriction for $b\to d\gamma$ 
decay rate) the decay asymmetry minimum and maximum
values are given.
We note, that for some values of $tan2\theta$,  $M_{W_R}$, $M_{\varphi^+}$ the decay asymmetry $a_{CP}$ from (8)
becomes abnomally large. This means, that for such values of
$tan2\theta$,  $M_{W_R}$, $M_{\varphi^+}$ (8) becomes incorrect
(imaginary part of the amplitude becomes non-small 
in comparison with the real part and it is necessary to take into account  more terms of perturbation theory on $\alpha_s$).
Nevertheless, it is reasonably safe to suggest that in this case the 
difference from the standard model predictions for
$M_{W_R},M_{\varphi^+}\leq 10TeV$ can be significant.
%We don't give in Fig 3 values of $|a_{CP}|$ which exceed 30\% as in this
%case formula (8) is not applicable (imaginary part of the amplitude
%becomes non-small in compare with the real part).\\
Thus, for the case of non-manifest left-right symmetry for the large
parameter space of the $SU(2)_L\times SU(2)_R \times U(1)$ model
($M_{W_R},M_{\varphi^+}\leq (10-15)TeV$, $\tan2\theta \geq (1.5-2.5)$)
one can expect a significant deviations from the standard model
predictions for $a_{CP}$ for both  decays.\\
Let us proceed to the two Higgs doublet extension of the standard model.
In general case Yukawa interaction of quarks with Higgs doublets
$\varphi_1$ and $\varphi_2$ is:
\begin{equation}
L=\bar{q}_L(\gamma_1^d\varphi_1+\gamma_2^d\varphi_2)d_R+             % (12)
\bar{q}(\gamma_1^u\varphi_1+\gamma_2^u\varphi_2)u_R,
\end{equation}
where $q_L$ is the quark doublet and $d_R$ and $u_R$ are quark singlets
and $\gamma_1^u$, $\gamma_2^u$, $\gamma_1^d$, $\gamma_2^d$
are matrices in flavor space \cite{15}.
Usually two versions of this model are considered
\cite{15}:\\
model I, where only one doublet ($\varphi_1$) interacts with quarks:
$\gamma_2^u=\gamma_2^d=0.$,\\
model II, where one of doublets interacts with up-type quarks and the
second one interacts with down-type quarks:
$\gamma_2^u=\gamma_1^d=0.$\\
In paper \cite{15} the model was considered where both
Higgs doublets interact with up and down quarks and
all of the quantities
$\gamma_1^u$, $\gamma_2^u$, $\gamma_1^d$, $\gamma_2^d$ are non-zero.\\
Generally speaking, in this case the flavor changing neutral currents can
arise \cite{15}.  The restrictions on Higgs particles masses and other
parameters in such a model were considered in \cite{21}.\\
The last  model (model III) is close in some respect to the
$SU(2)_{L}\times SU(2)_{R} \times U(1)$ model: for this  model, as for
the previous one, new CP-violating phase arises.
As for models I and II, there are no new sources of CP violation. \\
Formula for CP asymmetry in $b \to s \gamma$ decay  for two Higgs doublet
extension of the standard model is the following \cite{15}:
\begin{eqnarray}
\nonumber
a_{CP}(b\to s\gamma)  =
& - & 2\frac{2}{9}
\frac{(C^H_{s\gamma}C^{WH}_{sg}-C^{WH}_{s\gamma}C^{H}_{sg})\alpha_s}
{C^2_{s\gamma}}Im(\xi_t \xi_b)\\                                    % (13)
& -&2\frac{20}{81} \frac{Re(v_t^*v_u)+rRe(v_t^*v_c)}{|v_t|^2}
\frac{C^H_{s\gamma}c_1\alpha_s}{2C^2_{s\gamma}}Im(\xi_t \xi_b)\\
\nonumber
& - &2\frac{20}{81}\frac{(1-r)Im(v_t^*v_u)}{|v_t|^2}
\frac{(C^{WH}_{s\gamma}+Re(\xi_t \xi_b) C^H_{s\gamma})c_1\alpha_s}
{2C_{s\gamma}^2}
\end{eqnarray}
where $\xi_{t}$, $\xi_{b}$, $C^H_{s\gamma}$,$C^{WH}$,
$C^{WH}_{s\gamma}$, $C^{H}_{sg}$, $C_{s\gamma}$ are given in \cite{15}.
We note, that there is a difference between formula  (13)
and the expression for CP asymmetry in \cite{15}: in \cite{15} the factor
2/9 is missed. \\
In Table 3 the numerical results for the model III for some
values of charged Higgs boson masses are given. 
Generally speaking, values of
$a_{CP}$ for $b \rightarrow s \gamma$ decay,
given by Table 3, are lower than the results \cite{15} for the reason,
mentioned above, but the deviation from the standard model predictions
for relatively low masses of charged Higgs boson $\leq$ 200GeV (this is
within the limits given in \cite{21}) can be very large (more than 5
times).
As for the $b \rightarrow d \gamma$ decay (as it follows from the
Table 3 a)) the restrictions on absolute value $a_{CP}$ are close to 
the predictions of the standard model. The difference is that CP-asymmetry
here can have opposite sign and the minimum value of $a_{CP}$ can be very small.
In Table 3 b) the minimum and maximum values of CP asymmetry for
$b \rightarrow d \gamma$ decay, for the case when we use the 
less severe condition $\Delta < 50\%$ for the decay rate, are given. 
In this case the absolute
value of CP asymmetry can be 1.5 times larger than the standard model
predictions.\\
As we have mentioned above,  for the models I and II there is no
new source of CP-violation and as it follows from the Tables 4 and 5,
the values of $a_{CP}$ for two decays are almost the same as for the 
standard model.\\
In conclusion, the direct CP-violation in
$SU(2)_{L}\times SU(2)_{R} \times U(1)$
model and two-Higgs doublet extension of the standard model for
$b \rightarrow d\gamma$ and $b \rightarrow s\gamma$
decays  was investigated. The calculated values of  CP-asymmetry differ
from the standard model predictions and can have a sign opposite to that of
the latter. The difference is much stronger for non-manifest left-right 
symmetric model and  two-Higgs doublet extension of the standard
model (model III). \\
Authors want to thank A. Ali for stimulating discussions. One of
the authors (H. A.) wants to thank High Energy Group of ICTP for
hospitality.\\
The research described in this publication was made possible in part due
to the contract INTAS-93-1630.

%\vspace{0.5cm}
\newpage
\begin{center}
Figure Captions
\end{center}
\vspace{0.3cm}

Fig. 1. Maximum and minimum values of $a_{CP}$ in
$SU(2)_L \times SU(2)_R \times U(1)$ model for the case (1) and
a) for $b \rightarrow d\gamma$ decay, b) for $b \rightarrow s\gamma$ decay
for different values of $M_{W_{R}}$ and $M_{\varphi^{+}}$:
$M_{W_{R}}=M_{\varphi^{+}}$=5TeV
(curves 1 and 2); $M_{W_{R}}=M_{\varphi^{+}}$=10TeV
(curves 3 and 4);
$M_{W_{R}}=M_{\varphi^{+}}$=20TeV (curves 5 and 6);
$M_{W_{R}}=M_{\varphi^{+}}$=50TeV (curves 7 and 8).

\vspace{0.3cm}

Fig. 2. Maximum and minimum values of $a_{CP}$ in
$SU(2)_L \times SU(2)_R \times U(1)$ model for the case (2a) and
a) for $b \rightarrow d\gamma$ decay, b) for $b \rightarrow s\gamma$ decay
for different values of $M_{W_{R}}$ and $M_{\varphi^{+}}$:
$M_{W_{R}}=M_{\varphi^{+}}$=5TeV
(curves 1 and 2); $M_{W_{R}}=M_{\varphi^{+}}$=10TeV
(curves 3 and 4);
$M_{W_{R}}=M_{\varphi^{+}}$=20TeV (curves 5 and 6);
$M_{W_{R}}=M_{\varphi^{+}}$=50TeV (curves 7 and 8).

\vspace{0.3cm}

Fig. 3. Maximum and minimum values of $a_{CP}$ in
$SU(2)_L \times SU(2)_R \times U(1)$ model for the case (2b) and
for $b \rightarrow d\gamma$ decay
for different values of $M_{W_{R}}$ and $M_{\varphi^{+}}$:
$M_{W_{R}}=M_{\varphi^{+}}$=5TeV
(curves 1 and 2); $M_{W_{R}}=M_{\varphi^{+}}$=10TeV
(curves 3 and 4);
$M_{W_{R}}=M_{\varphi^{+}}$=20TeV (curves 5 and 6);
$M_{W_{R}}=M_{\varphi^{+}}$=50TeV (curves 7 and 8).

\newpage
\setcounter{page}{13}
\begin{center}
Table 1: Minimum and maximum values of $a_{CP}$ for case (1) for $b \to d \gamma$ and  $b\to s \gamma$ decays for $tan2\theta= 2.0$,
$M_{W_{R}}=M_{\varphi^{+}}$=10TeV and for various values of $\alpha$.
\end{center}
\vspace{0.1cm}

\begin{tabular}{|c|c|c|}  \hline
 $\alpha$ & $a_{CP}(b\to d\gamma)$ &
$a_{CP}(b\to s\gamma)$ \\  \hline
-2.2 & $0.043\div 0.107$ & $-0.0101 \div -0.0045$ \\  \hline
-2.0 & $0.041\div 0.130$ &$-0.096 \div -0.0040$ \\  \hline
-1.8 & $ 0.049\div 0.123$ & $-0.0087 \div -0.0039$  \\ \hline
 1.8 & $0.0085 \div 0.162$ & $-0.0037 \div 0.0019$ \\  \hline
 2.0 & $-0.0024 \div 0.172$ & $ -0.0039\div 0.0039$ \\ \hline
 2.2 & $-0.0027\div 0.160$ & $-0.0018 \div 0.0044$  \\  \hline
\end{tabular}
\vspace{0.5cm}

\begin{center}
Table 2: Minimum and maximum values of $a_{CP}$ for case (2a) for $b \to d \gamma$ and $b\to s \gamma$ decays for $tan2\theta= 3.0$,
$M_{W_{R}}=M_{\varphi^{+}}$=10TeV and for various values of $\alpha$.
\end{center}
\vspace{0.1cm}

\begin{tabular}{|c|c|c|}  \hline
 $\alpha$ & $a_{CP}(b\to d\gamma)$ &
$a_{CP}(b\to s\gamma)$ \\  \hline
-3.0 & $-0.017\div 0.044$ & $-0.0033 \div -0.0045$ \\  \hline
-2.5 & $-0.035\div 0.062$ &$-0.0114 \div -0.0026$ \\  \hline
-2.0 & $ 0.008\div 0.110$ & $-0.0145 \div -0.0042$  \\ \hline
 2.0 & $-0.031 \div 0.188$ & $-0.0134 \div 0.0102$ \\  \hline
 2.5 & $-0.036 \div 0.108$ & $ -0.0008\div 0.0104$ \\ \hline
 3.0 & $-0.012\div 0.055$ & $-0.0018 \div 0.0028$  \\  \hline
\end{tabular}
\vspace{0.5cm}

\newpage

\begin{center}
Table 3: Values of CP-asymmetry for $b \rightarrow s + \gamma$
and $b \rightarrow d + \gamma$  for model III.
a) for $\Delta < 10\%$,
b) for $\Delta < 50\%$
\end{center}
\vspace{0.1cm}

a)

\begin{tabular}{|c|c|c|}  \hline
 $m_{H^{+}}, GeV$ &
$-a_{CP}(b\to s\gamma)/10^{-3}$ & $a_{CP}(b\to d\gamma)/10^{-2}$
\\  \hline
50 & $-53 \div 53$ & $-17 \div 17$ \\  \hline
100 & $-45 \div 45$ & $-17 \div 17$ \\  \hline
200 & $-35 \div 35$ & $-18 \div 18$ \\  \hline
400 & $-25 \div 25$ & $-18 \div 18$ \\  \hline
800 & $-16 \div 16$ & $-18 \div 18$ \\  \hline
1600 & $-9.7 \div 9.5$ & $-17 \div 18$ \\  \hline
3200 & $0.6 \div 7$ & $0.6 \div 17$ \\  \hline
6400 & $2 \div 7$ & $3.1\div 17$ \\  \hline
12800 & $2.6 \div 6.8$ & $3.5 \div 17$ \\  \hline
25600 & $2.8 \div 6.5$ & $3.6 \div 16$ \\  \hline
\end{tabular}
\vspace{0.5cm}

b)

\begin{tabular}{|c|c|c|}  \hline
 $m_{H^{+}}, GeV$ &
$-a_{CP}(b \to s\gamma)/10^{-3}$ & $a_{CP}(b \to d\gamma)/10^{-2}$
\\  \hline
50 & $-71 \div 71$ & $-23 \div 23$ \\  \hline
100 & $-61 \div 61$ & $-23 \div 23$ \\  \hline
200 & $-48 \div 48$ & $-23 \div 23$ \\  \hline
400 & $-34 \div 34$ & $-24 \div 24$ \\  \hline
800 & $-22 \div 21$ & $-24 \div 24$ \\  \hline
1600 & $-12 \div 12$ & $-22 \div 25$ \\  \hline
3200 & $-5 \div 11$ & $-4.1 \div 25$ \\  \hline
6400 & $0.4\div 8.3$ & $2.4\div 20$ \\  \hline
12800 & $2.3 \div 6.8$ & $3.5 \div 17$ \\  \hline
25600 & $2.8 \div 6.5$ & $3.6 \div 16$ \\  \hline
\end{tabular}
\newpage
\begin{center}
Table 4: Values of CP-asymmetry for the Model I\\
a) for $\Delta < 10\%$, b) for $\Delta < 50\%$
\end{center}

a)

\begin{tabular}{|c|c|c|}  \hline
 $m_{H^{+}}, GeV$ &
$-a_{CP}(b \to s \gamma)/10^{-3}$ & $a_{CP}(b\to d \gamma)/10^{-2}$
\\  \hline
50 & $2.9 \div 6.7$ & $3.6 \div 17$ \\  \hline
100 & $2.9 \div 6.7$ & $3.6 \div 17$ \\  \hline
200 & $2.9 \div 6.7$ & $3.6 \div 17$ \\  \hline
800 & $2.9 \div 6.7$ & $3.6 \div 17$ \\  \hline
3200 & $2.9 \div 6.5$ & $3.6 \div 16$ \\  \hline
\end{tabular}
\vspace{0.2cm}

b)

\begin{tabular}{|c|c|c|}  \hline
 $m_{H^{+}}, GeV$ &
$-a_{CP}(b\to s\gamma)/10^{-3}$ & $a_{CP}(b\to d\gamma)/10^{-2}$
\\  \hline
50 & $2.9 \div 8.6$ & $3.6 \div 21$ \\  \hline
100 & $2.9 \div 8.9$ & $3.6 \div 22$ \\  \hline
200 & $2.9 \div 8.7$ & $3.6 \div 22$ \\  \hline
800 & $2.9 \div 7.6$ & $3.6 \div 19$ \\  \hline
3200 & $2.9 \div 6.5$ & $3.6 \div 17$ \\  \hline
\end{tabular}
\vspace{0.5cm}
\begin{center}
Table 5: Values of CP-asymmetry for  the Model 2\\
a) for $\Delta < 10\%$, b) for $\Delta < 50\%$
\end{center}

a)

\begin{tabular}{|c|c|c|}  \hline
 $m_{H^{+}}, GeV$ &
$-a_{CP}(b\to s\gamma)/10^{-3}$ & $a_{CP}(b\to d\gamma)/10^{-2}$
\\  \hline
400 & - & -\\  \hline
650 & - & - \\  \hline
1300 & $2.8 \div 6.1$ & $3.4 \div 15$ \\  \hline
2600 & $2.8 \div 6.2$ & $3.5 \div 16$ \\  \hline
5200 & $2.9 \div 6.3$ & $3.6 \div 16$ \\  \hline
\end{tabular}

\vspace{0.2cm}

b)

\begin{tabular}{|c|c|c|}  \hline
 $m_{H^{+}}, GeV$ &
$-a_{CP}(b \to s \gamma)/10^{-3}$ & $a_{CP}(b\to d\gamma)/10^{-2}$
\\  \hline
400 & $2.4 \div 5.3$ & $2.9 \div 13$ \\  \hline
650 & $2.5 \div 5.7$ & $3.1 \div 14$ \\  \hline
1300 & $2.7 \div 6.1$ & $3.4 \div 15$ \\  \hline
2600 & $2.8 \div  6.3$ & $3.5 \div 16$ \\  \hline
5200 & $2.9 \div 6.3$ & $3.6 \div 16$ \\  \hline
\end{tabular}
\vspace{0.5cm}

\newpage
\begin{figure}[htb]
\epsfxsize=11cm
\epsfysize=6.4cm
\mbox{\hskip 0.8in}\epsfbox{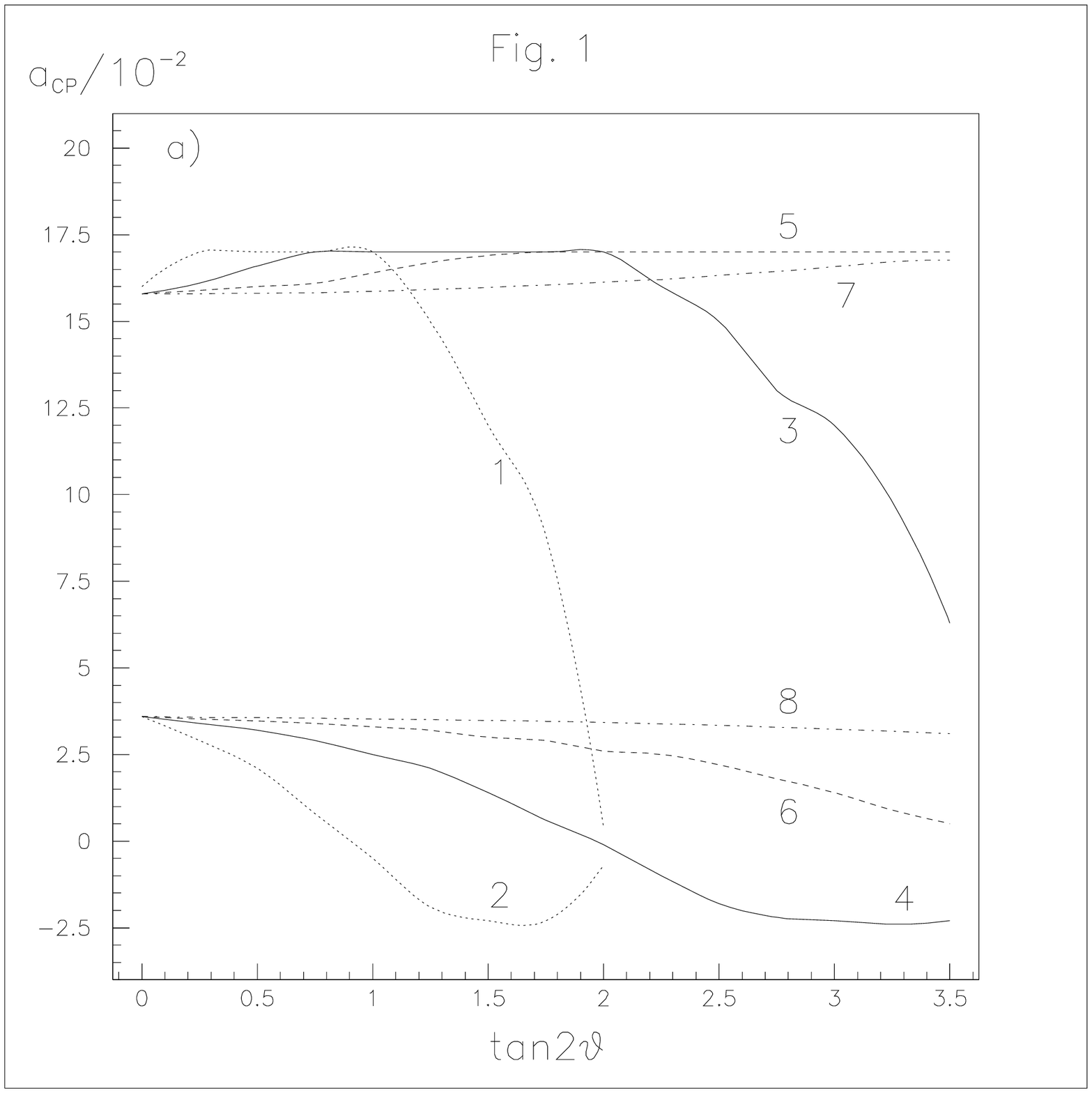}
\end{figure}
\begin{figure}[htb]
\epsfxsize=11cm
\epsfysize=6.4cm
\mbox{\hskip 0.8in}\epsfbox{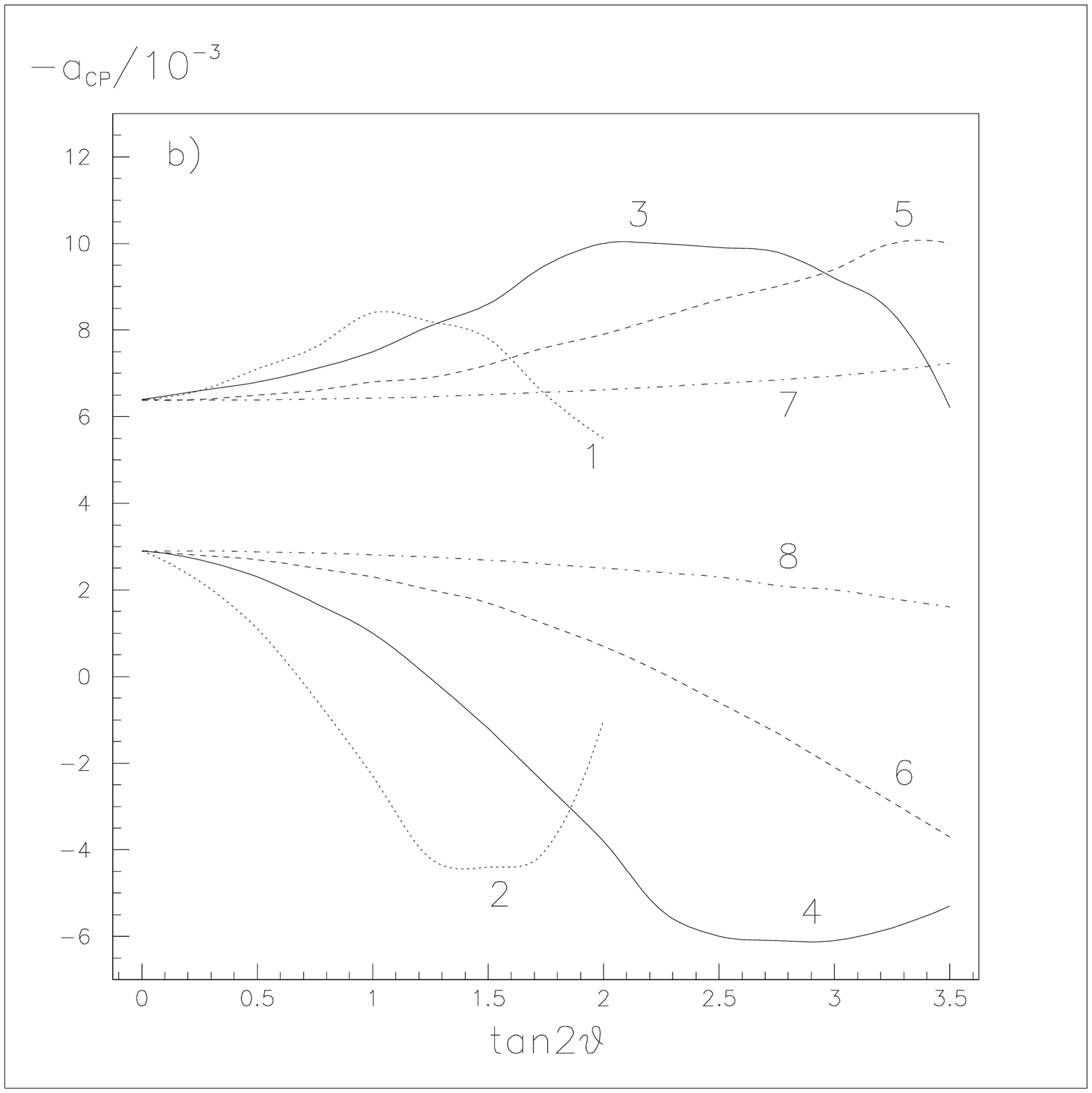}
\end{figure}
\newpage
\begin{figure}[htb]
\epsfxsize=11cm
\epsfysize=6.4cm
\mbox{\hskip 0.8in}\epsfbox{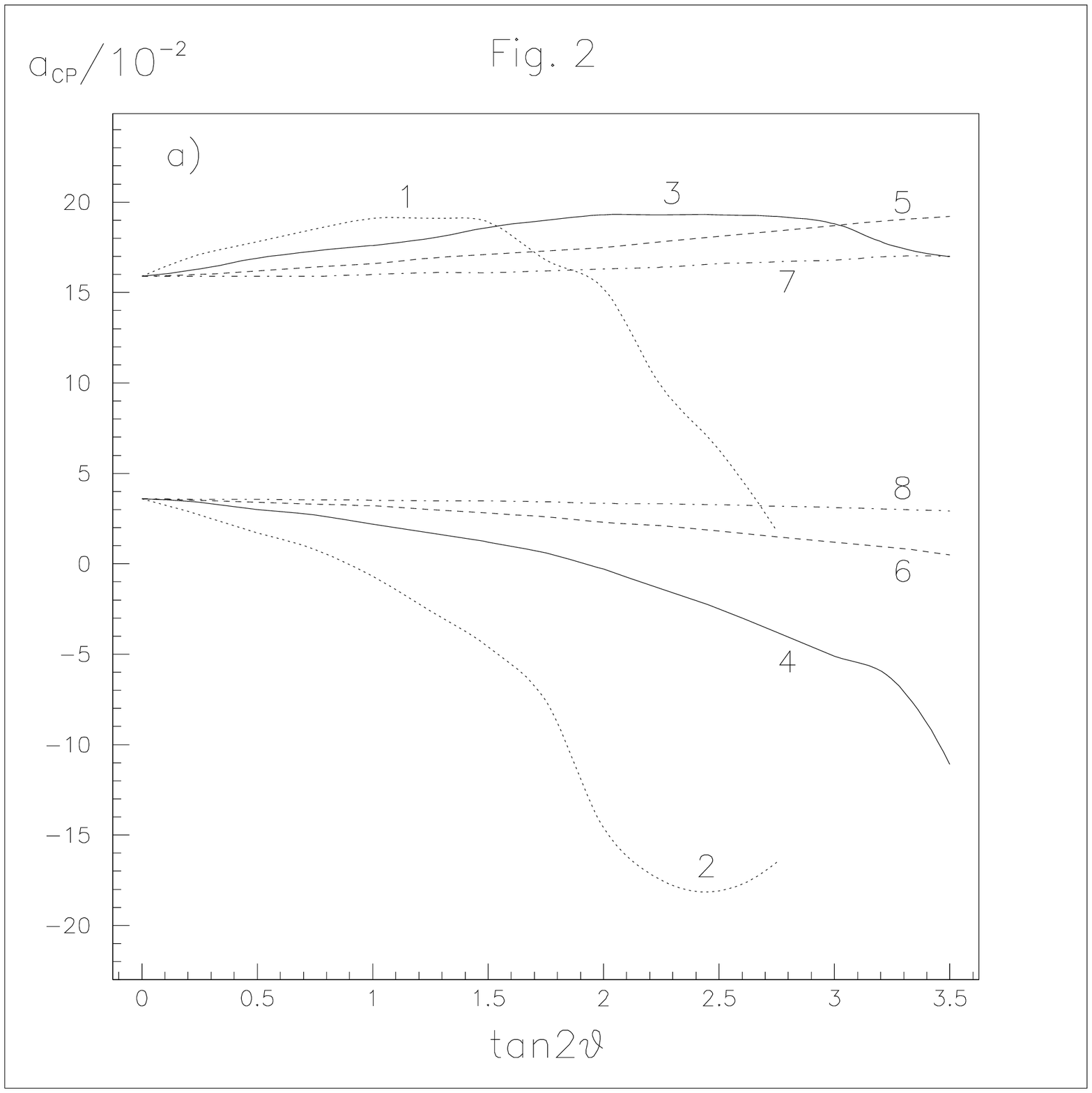}
\end{figure}
\begin{figure}[htb]
\epsfxsize=11cm
\epsfysize=6.4cm
\mbox{\hskip 0.8in}\epsfbox{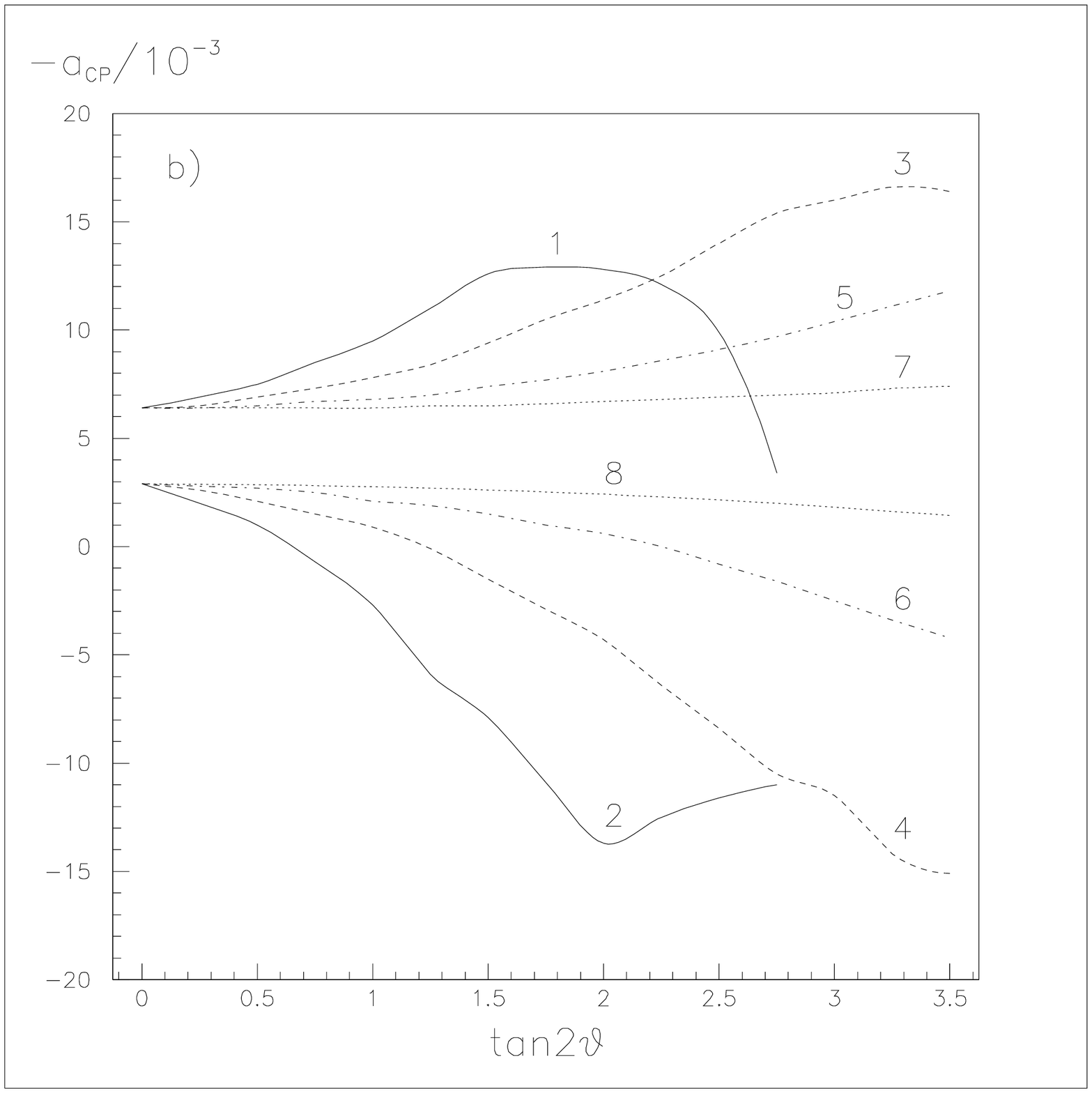}
\end{figure}
\newpage
\begin{figure}[htb]
\epsfxsize=11cm
\epsfysize=15cm
\mbox{\hskip 0.8in}\epsfbox{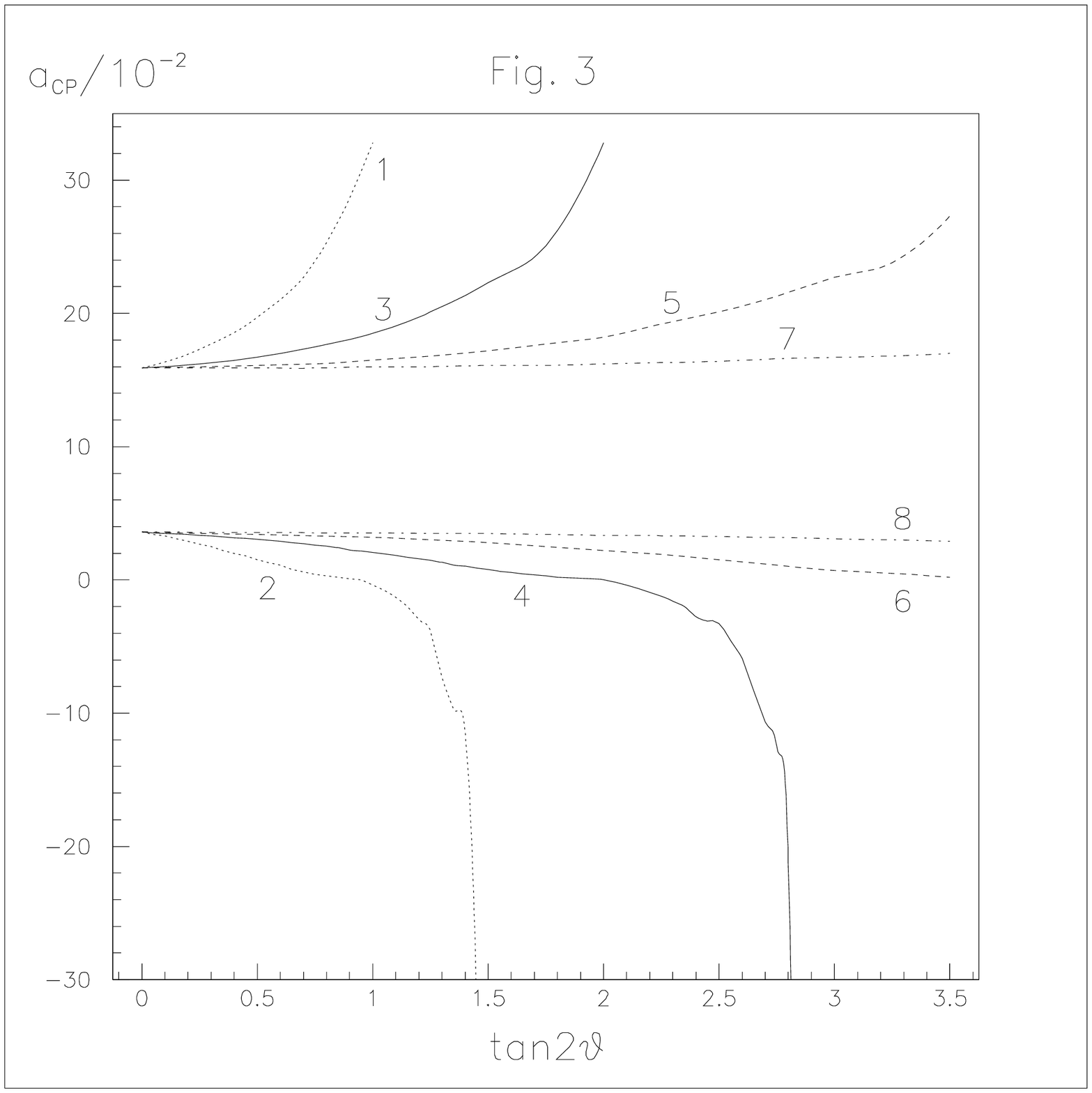}
\end{figure}


\vspace{0.5cm}
\begin{thebibliography}{99}
\bibitem{1} R.Ammar et al. (CLEO Collaboration), Phys.Rev.Lett. 71 (1993)
674.
\bibitem{2} M. S. Alam et al. (CLEO Collaboration), Phys.Rev.Lett. 74 (1995)
2885.
\bibitem{3} B. Grinstein and M. B. Wise, Phys. Lett. B201, (1988), 274.
\bibitem{4} W. S. Hou, R. S. Willey, Phys. Lett. B202, (1988), 591.
\bibitem{5}R. Barbieri and G. F. Guidice, Phys. Lett. B309, (1993), 86.
\bibitem{6} S. Bertolini et al. Nucl. Phys. B353, (1991), 591.
\bibitem{7} L. Randal and R Sundrum, MIT preprint MIT-CTP-2211 (1993).
\bibitem{8} D.Cocolicchio et al, Phys.Rev. D40 (1989) 1477.
\bibitem{9} H.M.Asatrian and A.N.Ioannissian, Mod.Phys.Lett.A5 (1990) 1089;
Sov. Journ. of Nucl. Phys. 51 (1990) 858.
\bibitem{10} K.S.Babu, K.Fujikawa and A.Yamada,Phys.Lett. B333 (1994) 196.
\bibitem{11} P.Cho and M.Misiak, Phys.Rev. D49 (1994) 5894.
\bibitem{12} C. Greub, T. Hurth, D. Wyler Phys. Rev. D54 (1996) 3350.
\bibitem{13} G. Ecker and W. Grimus, in: Moriond 86, v.1, p. 201.
\bibitem{14} J.Soares, Nucl.Phys.B 367 (1991) 575.
\bibitem{15} L.Wolfenstein and Y.L.Wu, Phys.Rev.Lett. 73 (1994) 2809.
\bibitem{16} H. M. Asatrian, A. N. Ioannissian, Phys. Rev. D54 (1996)
5242.
\bibitem{17} A. Ali DESY Report No. 96-106, HEP-PH/9606324.
\bibitem{18} R.N.Mohapatra, in "CP Violation" ed. by C.Jarslkog,World
Scientific, page 384.
\bibitem{19} M. Bando et al., Mod. Phys. Lett. A7, 3379 (1992)
\bibitem{20} A. Ali, D. London. DESY Report No.96-140, HEP-PH/9607392.
\bibitem{21} D. Atwood, L. Reina, A. Soni. CEBAF preprint, JL-TH-96-15,
hep-ph/9609279.
\end{thebibliography}
\end{document}